\documentclass[aps,prl,twocolumn,showpacs,groupedaddress]{revtex4}
\newcommand{\Tr}{\text{Tr}}
\usepackage{graphicx}
\begin{document}
\title{Entanglement induced in spin-$\frac 1 2$ particles by a spin-chain near
its critical points}
\author{X.X. Yi, H. T. Cui, and L. C. Wang}
\affiliation{Department of Physics, Dalian University of
Technology, Dalian 116024, China}
\date{\today}
\begin{abstract}
A relation between entanglement and criticality of spin chains is
established. The entanglement we exploit is shared between
auxiliary particles, which are isolated from each other, but are
coupled to the same critical spin-$1/2$ chain. We analytically
evaluate the reduced density matrix, and numerically show the
entanglement of the auxiliary particles in the proximity of the
critical points of the spin chain. We find that the entanglement
induced by the spin-chain may reach one, and it can signal very
well the critical points of the   chain. A physical understanding
and experimental  realization with trapped ions are presented.
\end{abstract}
 \maketitle

Quantum entanglement lies at the heart of the difference between
the quantum and classical multi-particle world, and  can be
treated as a useful resource in various tasks such as
cryptography, quantum computation and teleportation
\cite{nielsen00}. Quantum phase transitions\cite{sachdev99} are
transitions between quantitatively distinct phases of quantum
many-body systems, driven solely by quantum fluctuations. In the
past decade, a great effort has been devoted to understand the
relations between entanglement and quantum phase transitions
\cite{osborne02, vidal03, osterloh02, wu04, li04,gu04,
verstraete04}. In fact, it is natural to associate the quantum
phase transition and entanglement once correlations are behind
both of them. By sharing this point of view, one anticipates that
entanglement induced by a quantum critical many-body system  will
furnish a dramatic signature of quantum critical points for the
many-body system.

On the other hand,  we usually think of environment that surrounds
quantum system  as a source of decoherence.   Recently researchers
have started to investigate the positive effects\cite{plenio99,
bose99, beige00, horodecki01, braun02, yi03, beuatti03, shresta05}
of environment, for example, environment assisted information
processing and environment induced  entanglement.  These
investigations pave a new way to engineer mechanisms of
preventing, minimizing or using the impact of environment in
quantum information processing.  In those works, however, the
environment was modelled as a set of independent quantum systems,
i.e., correlations among particles in the environment were
ignored. An interesting open question is whether the correlation
among environmental particles can affect the entanglement induced
in a bipartite system that couples to it.

In this paper, we show how to exploit  entanglement in auxiliary
particles induced by a quantum critical  many-body system  as an
essential tool to reveal quantum phenomena in the many-body
quantum system. Indeed, quantum phase transitions are accompanied
by a qualitative change in the nature of classical correlations,
such drastic changes in the properties of ground states are often
due to the collectiveness/randomness of interparticle couplings
which are possibly reflected in entanglement between systems that
couple to it. Here we adopt a spin-chain system described by the
one-dimensional spin-$\frac 1 2 $ XY model as the many-body
system. Another pair of spin-$\frac 1 2 $ systems that couple to
the spin chain would act as the auxiliary particles.  We observe
that the entanglement of the auxiliary particles changes sharply
in the proximity of quantum phase transition. This change can be
traced down to the presence of collectiveness in the dominant
couplings of the auxiliary particles to the chain, then it
reflects the critical properties of the many-body system. This
observation offers a new tool to study quantum critical phenomena,
in particular for more general systems, where analytical solutions
might not be available. A possible realization of this scheme is
proposed with trapped ions. It utilizes off-resonant
standing-waves driven ions in traps\cite{duan03, porras04} to
simulate the one-dimensional XY spin-chain. The proposal also
could be realized with ultracold atoms superposed by optical
lattice\cite{pachos04}. This makes the study appealing for
experimental test.

Consider a bipartite system consisting of two spin-$\frac 1 2 $
particles $a$ and $b$, and a quantum many-body system described by
the one dimensional spin-$\frac 1 2 $ XY model. The system
Hamiltonian $H_S$ and the Hamiltonian $H_B$ of the chain read,
\begin{eqnarray}
H_S&=&\frac{\omega_a}{2}\sigma_a^z+\frac{\omega_b}{2}\sigma_b^z,\nonumber\\
H_B&=&-\sum_{l=1}^N(\frac{1+\gamma}{2}\sigma_l^x\sigma_{l+1}^x+\frac{1-\gamma}{2}\sigma_l^y\sigma_{l+1}^y
+\frac{\lambda}{2}\sigma_l^z),\nonumber\\
\end{eqnarray}
where $N$ is the number of sites, $\sigma_i^{\alpha}
(\alpha=x,y,z)$ are the pauli matrices, and $\gamma$ is the
anisotropy parameter. The periodic boundary condition
$\sigma_{N+1}=\sigma_1$ is assumed for the spin chain. Suppose the
coupling of  the auxiliary particles to chain take the form
\begin{equation}
H_I=\sum_{l=1}^N(g\sigma_a^z\sigma_l^z+h\sigma_b^z\sigma_l^z),
\end{equation}
where $g$ and $h$ denote coupling constants. Clearly, $[H_S,
H_I]=0,$ which implies that the energy of the auxiliary particles
is conserved, but the coherence may  not be conserved depending on
the detail of the system-chain coupling.   This leads to the
following form of the time evolution operator $U(t)$,  $
U(t)=\sum_{i,j=0,1}U_{ij}(t)|ij\rangle\langle ij|, $ where
$|ij\rangle=|i\rangle_a\otimes |j\rangle_b,$ and $|i\rangle_a
(|i\rangle_b, i=0,1)$ represent the eigenstates of
$\sigma_a^z(\sigma_b^z)$. It is easy to show that $U_{ij}(t)
(i,j=0,1)$ satisfy $i\hbar\frac{\partial}{\partial
t}U_{ij}(t)=H_{ij}U_{ij}(t)$ with $
H_{ij}=-\sum_{l=1}^N(\frac{1+\gamma}{2}\sigma_l^x\sigma_{l+1}^x+\frac{1-\gamma}{2}\sigma_l^y\sigma_{l+1}^y
+\frac{\Lambda_{ij}}{2}\sigma_l^z), $ where
$\Lambda_{ij}=\lambda+(-1)^{i+1}2g+(-1)^{j+1}2h$, $i,j=0,1.$ If
the auxiliary particles are initially in state $|ij\rangle$, the
dynamics and statistical properties of the spin chain would be
govern by $H_{ij}$, it takes the same form as $H_B$ but with
modified field strengths $\Lambda_{ij}$. The Hamiltonian $H_{ij}$
can be diagonalized by a standard procedure\cite{lieb61} to be $
H_{ij}=\sum_k\omega_{ij,k}(\eta_{ij,k}^{\dagger}\eta_{ij,k}-\frac
1 2 ), $ where $\eta_{ij,k} (\eta_{ij,k}^{\dagger})$ are the
annihilation (creation) operators of the fermionic modes with
frequency
$\omega_{ij,k}=\sqrt{\varepsilon_{ij,k}^2+\gamma^2\sin^2\frac{2\pi
k}{N}}, $  $\varepsilon_{ij,k}=(\cos\frac{2\pi
k}{N}-\Lambda_{ij}), $ $ k=-N/2, -N/2+1,...,N/2-1.$ The fermionic
operator $\eta_{ij,k}$ was defined by the Bogoliubov
transformation as, $
\eta_{ij,k}=d_k\cos\frac{\theta_{ij,k}}{2}-id_{-k}^{\dagger}
\sin\frac{\theta_{ij,k}}{2},
$ where $d_k=\frac{1}{\sqrt{N}}\sum_la_l \mbox{exp}(-i2\pi lk/N),$
and the mixing angle $\theta_{ij,k}$ was defined by
$\cos\theta_{ij,k}=\varepsilon_{ij,k}/\omega_{ij,k}.$ Fermionic
operators $a_l$ were connected with the spin operators by the
Jordan-Wigner transformation via
$a_l=(\prod_{m<l}\sigma_m^z)(\sigma_l^x+i\sigma_l^y)/2$. The
  operators $\eta_{ij,k}$  parameterized by $i$ and
$j$ clearly do not commute with each other, this will leads to
entanglement in the auxiliary particles as shown later on. Before
going on to calculate the reduced density matrix, we present a
discussion on the diagonalization of  $H_{ij}$. For a chain with
periodic boundary condition, i.e., $\sigma_1=\sigma_N,$ boundary
terms $H_{boun}\sim [(a^{\dagger}_Na_1+\gamma
a_Na_1)+h.c.][exp(i\pi M)+1]$ have to be taken into
acount\cite{lieb61, katsura62}.   In this paper, we would work
with $H_{boun}$ ignored\cite{note1}, because we are interested in
finding a link between the criticality of the chain and the
entanglement in the auxiliary particles.

Having given an initial product(separable) state of the total
system, $|\psi(0)\rangle=$ $
|\phi_a(0)\rangle\otimes|\phi_b(0)\rangle$ $
\otimes|\phi_B(0)\rangle,$ we can obtain the reduced density
matrix for the auxiliary particles as $\rho_{ab}(t)=$ $
\Tr_B[U(t)|\psi(0)\rangle\langle\psi(0)|U^{\dagger}(t)],$ it may
be formally written in the form
\begin{equation}
\rho_{ab}(t)=\sum_{i,j,m,n}\rho_{ij;mn}(t)|ij\rangle\langle
mn|.\label{stateab0}
\end{equation}
A straightforward but somewhat tedious calculation shows that
\begin{widetext}
\begin{eqnarray}
\rho_{ij;mn}(t)&=&\rho_{ij;mn}(t=0)\Gamma_{ij;mn}(t),\nonumber\\
\Gamma_{ij;mn}(t)&=&\prod_ke^{\frac i 2
(\omega_{ij,k}-\omega_{mn,k})t}
\{1-(1-e^{i\omega_{ij,k}t})\sin^2\frac{\theta_k-\theta_{ij,k}}{2}-
(1-e^{-i\omega_{mn,k}t})\sin^2\frac{\theta_k-\theta_{mn,k}}{2}\nonumber\\
&+&(1-e^{i\omega_{ij,k}t})
(1-e^{-i\omega_{mn,k}t})[\sin\frac{\theta_k-\theta_{ij,k}}{2}\sin\frac{\theta_k-\theta_{mn,k}}{2}
\cos\frac{\theta_{ij,k}-\theta_{mn,k}}{2}]\}, \label{stateab}
\end{eqnarray}
\end{widetext}
where $\cos\theta_k=\frac{\cos(2\pi k/N)-\lambda}{\sqrt{(\cos
(2\pi k/N)-\lambda)^2+\gamma^2\sin^2(2\pi k/N)}}.$ To derive this
result, the spin chain was assumed to be initially in the ground
state of $H_B$. Discussions on Eq.(\ref{stateab}) are in order.
For $g=h$, one has $\Lambda_{ij}=\lambda$ when $i=0$ and $j=1$, or
$i=1$ and $j=0.$ Hence, $|01\rangle$ and $|10\rangle$ expand a
decoherence free subspace. Subsequently, the entanglement of state
in this subspace remains unchanged due to $\Gamma_{ij;mn}(t)=1.$
The situation changes if $h\neq g$, where the decoherence free
subspace does not exist. The entanglement shared between the
auxiliary particles evolves with time according to
Eq.(\ref{stateab}) in this situation.

With these expressions, we now turn to study entanglement shared
between the auxiliary particles $a$ and $b$ in state
(\ref{stateab0}). To be specific, we choose
$|\phi_a(0)\rangle\otimes|\phi_b(0)\rangle$
$=1/\sqrt{2}(|0\rangle_a+|1\rangle_a$ $\otimes
1/\sqrt{2}(|0\rangle_b+|1\rangle_b$ as the initial state of the
auxiliary particles, while the spin chain is assumed to be in the
ground state of Hamiltonian $H_B$. The entanglement measured by
the Wootters concurrence can be calculated and the numerical
result was shown in figures 1-4. The regions of criticality appear
when the ground and first excited states become degenerate. We
first focus on the criticality in the XX model. The XX model,
which corresponds to $\gamma=0$, has a criticality region along
the line between $\lambda=1$ and $\lambda=-1$. The criticality is
reflected in the entanglement of the auxiliary particles, which
appear in figure 1 and 2.
\begin{figure}
\includegraphics*[width=0.9\columnwidth,
height=0.8\columnwidth]{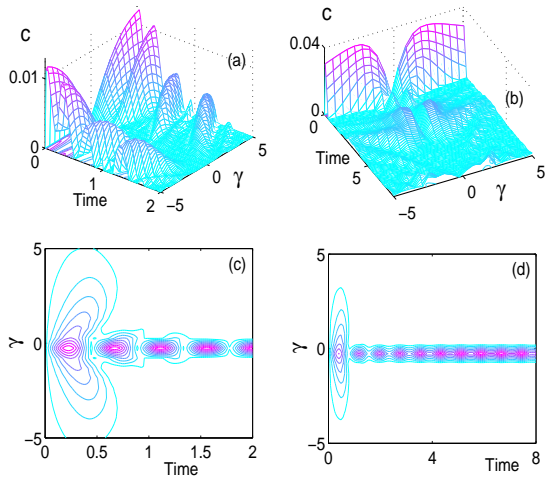} \caption{(color
online)Wootters concurrence of auxiliary particles as a function
of time and the anisotropy parameter $\gamma$. The figure was
plotted for $N=1502$ sites, and $g=h=0.05$. Figure
1-(a),(b),(c)and (d) correspond to $\lambda=5$,
$\lambda=2$,$\lambda=1$ and $\lambda=0.8$, respectively.}
\label{fig1}
\end{figure}
Figure 1 shows the Wootters concurrence as a function of time and
the anisotropy parameter $\gamma$, it is clear that the Wootters
concurrence has a sharp change in the limit $\gamma\rightarrow 0.$
This result can be understood by considering the value of
$\theta_{ij,k}$, which take $0$ or $\pi$ depending on the sign of
$\cos(2\pi k/N)-\Lambda_{ij}$ in this limit. In either case
$|\Gamma_{ij;mn}(t)|=1$, which indicates that the norm of any
element of the reduced density matrix remains unchanged.
\begin{figure}
\includegraphics*[width=0.9\columnwidth,
height=0.4\columnwidth]{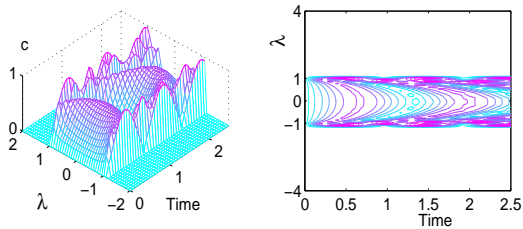} \caption{(color
online)Wootters concurrence {\it versus} time and $\lambda$ with
fixed $\gamma=10^{-4}$ (close to zero). This corresponds to the XX
model, and the critical points with $\lambda=\pm 1$ are clearly
shown. The figure on the left is a contour plot for the right. The
other parameters chosen are $N=1502$, and $g=h=0.05$. }
\label{fig2}
\end{figure}
Figure 2 shows the entanglement of the auxiliary particles near
the criticality region with $\gamma=0$ and $\lambda=\pm 1$. The
entanglement changes sharply along the line of $\lambda=\pm 1$ in
the time-$\lambda$ plane. This is a reflection of the critical
phenomena in the entanglement of the auxiliary particles.

\begin{figure}
\includegraphics*[width=0.9\columnwidth,
height=0.5\columnwidth]{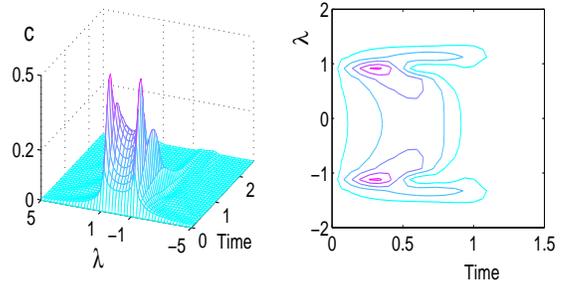} \caption{(color
online)Entanglement(measured by Wootters concurrence) induced by
the transverse Ising spin-$\frac 1 2 $ chain, which may be
obtained from the XY model by setting $\gamma=1$. The numerical
calculation was performed with $N=1502$ sites, and $g=h=0.05.$}
\label{fig3}
\end{figure}
Next we turn to study the criticality in the transverse Ising
model. The structure of the transverse model ground state changes
dramatically as the parameter $\lambda$ is varied. The dependence
of the entanglement on $\lambda$ is quite
complicated\cite{sachdev99}. Here we present analyses for the
$\lambda\rightarrow \infty$, $\lambda=1$ and $\lambda=0$ limits.
In the $\lambda\rightarrow \infty$ limit the ground state of the
chain approaches a product of spins pointing the positive $z$
direction. The mixing angle $\theta_{ij,k}$ and $\theta_k$ tend to
$-\pi$ in this limit. This leads to $\Gamma_{ij;mn}(t)=1$, which
indicates that the initial state does not evolve with time, i.e.,
the auxiliary particles remain in separable states. The
$\lambda=0$ limit is fundamentally different from the
$\lambda\rightarrow \infty$ limit because the corresponding ground
state is doubly degenerate under the global spin flip by
$\prod_{l=1}^N\sigma_l^z$.   This symmetry  breaks at
$|\lambda|=1$  and the chain develops a nonzero magnetization
$\langle\sigma^x\rangle\neq 0$ which grows as $\lambda$ is
decreased. This dramatic change in the ground state of the spin
chain can be found in the entanglement of auxiliary particles in
figure 3. In fact, the ground state of the XY models is very
complicated with many different regimes of
behavior\cite{barouch70,yang05}. With whatever $\gamma$, there is
a sharp change in the entanglement across the line
$|\lambda|=1$(figure 4). This signals the change in the ground
state of the spin chain from paramagnetic phase to the others.
Although the entanglement in both cases with $\gamma=0$ and
$\gamma=1$ shares the same properties along the line of
$|\lambda|=1$, i.e., it changes dramatically across this critical
region, the two case are quite different in its nature. The two
auxiliary particles never evolve in the case  of $\gamma=0$, while
the two will eventually evolve into the pointer states in another
case(except for $\lambda\rightarrow \infty$), hence no
entanglement share between the auxiliary particles in the later
case in the $t\rightarrow\infty$ and $N\rightarrow\infty$ limit.
\begin{figure}
\includegraphics*[width=0.9\columnwidth,
height=0.4\columnwidth]{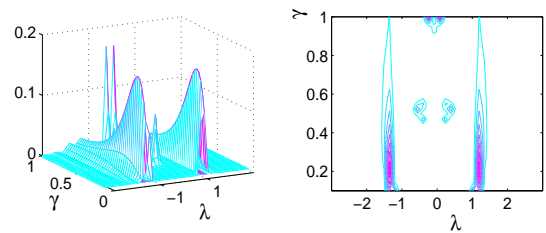} \caption{(color
online)Wootters concurrence at time $t=0.75$. The locations of
peaks on the $\lambda$ axis do not change with time. Various
parameters are $N=1502,$ $g=h=0.05$.} \label{fig4}
\end{figure}

The connection between the entanglement in auxiliary particles and
the criticality of spin chain may be understood as singularity in
one or more $\rho_{ij;mn}(t)$ (the elements of the reduced density
matrix) at the critical points. To show this point, we recall that
\begin{equation}
\rho_{ij;mn}(t)=\frac 1 4  \langle
\phi_B(0)|e^{-iH_{ij}t}e^{iH_{kl}t}|\phi_B(0)\rangle,
\end{equation}
with the same notations and initial states given  before
Eq.(\ref{stateab0}). Notice that $|\phi_B(0)\rangle$ was taken as
the ground state $|0\rangle_B$ of $H_B$, but it is not an
eigenstates of $H_{ij}$ with $\Lambda_{ij}\neq \lambda$. Expanding
$|0\rangle_B$ in terms of eigenstate $|\alpha\rangle_B^{ij}$ of
$H_{ij},$ $(\alpha=0,1,...,)$
\begin{equation}
|0\rangle_B=c_0^{ij}|0\rangle_B^{ij}+\sum_{\alpha\neq 0}
c_{\alpha}^{ij}|\alpha\rangle_B^{ij},
\end{equation}
one can easily prove that $|c_{\alpha\neq 0}^{ij}|$ $\sim h,g$
when $g\ll 1$ and $h\ll 1$. This is exactly the case under our
consideration. Where $|0\rangle_B^{ij}$ denotes the ground state
of $H_{ij}$. Therefore,
\begin{equation}
\rho_{ij;mn}(t)\sim  \frac 1 4 c_0^{ij} [c_0^{kl}]^*  \ \
^{kl}_B\langle 0|0\rangle_B^{ij}e^{-i(\Omega_{ij}-\Omega_{kl})t},
\end{equation}
up to first order in $g$ and $h$. Here $\Omega_{ij}$ stands for
the ground state energy of $H_{ij}$. As observed in
\cite{zanardi05}, a sudden drop in $^{kl}_B\langle
0|0\rangle_B^{ij}$ can signal the regions of criticality in the
spin chain. Therefore the entanglement as a function of
$\rho_{ij;mn}(t)$ can signal the criticality of the chain.

This discussion, apart from its theoretical interests, offers a
possible experimental method to study critical phenomena without
the need to identify the state of the system, in particular in the
presence of degeneracy. The XY model can be realized with trapped
ions under the action of off-resonant standing
waves\cite{porras04} as follows. Consider a system of trapped
ions, whose physical implementation corresponds to Coulomb chains
in Penning  traps or  an array of ion microtraps. The Coulomb
repulsion together with the trapped potential and ions' motion
yield a set of collective vibrational modes, these collective
modes can be coupled to the internal states of ions by conditional
forces. The directions of the conditional forces\cite{porras04}
would determine the effective couplings between the effective
spins. To simulate the XY model, we apply two conditional forces
in both directions $x$ and $y$. The effective coupling is
proportional to $1/r^3$, where $r$ represents the relative
distance between ions. Although the couplings are distance
dependent, the critical properties are shown to be similar to
these ideal models\cite{dutta01,porras04} considered in this
paper. The coupling of the auxiliary particles, which share
another trapping with the ions in the chain, to the chain can be
simulated in the same way, where a conditional force in the $z$
direction is applied. The particles in the chain may circle the
auxiliary particles, such that the couplings of the auxiliary
particles to the particles in the chain equal.  The proposal could
also  be realized with ultracold atoms in optical lattice by the
method represented in \cite{angelo05}. We would like to notice
that preventing any energy exchange between the chain and the
auxiliary particles might be a challenging task in experiments.
The requirement of uniform coupling also is experimentally very
challenging. Nevertheless, this paper established a link between
the entanglement and criticality from the other aspect, which
might shed light on the understanding of criticality from the
auxiliary particles, and it may open up a new way to characterize
and experimentally detect quantum phase transition from the
viewpoint of quantum information theory\cite{hartmann05}, where
two ancillas coupled to two particles at different sites in the
chain are considered. By this model, the relation between the
state transfer quality and the spectral gap of the chain was
established.

To conclude,  we have proposed a new method to study critical
phenomena in many-body systems. The criticality was found to have
a reflection in the entanglement of auxiliary particles that
couple to it.  Indeed, we have found that the entanglement change
dramatically along the line of critical points. This dramatic
change has been explained in terms of reflection of quantum phase
transitions, which lead to the collectiveness in the couplings of
the spin chain to the auxiliary particles. Quantum critical
many-body system  can also make a reflection for its criticality
in decoherence of a quantum system which couples to
it\cite{quan05}. In addition to shed light on the role of
environment in quantum information processing, these pave a new
way to study critical phenomena in many-body systems. The
generalization of these results to a wide variety of critical
phenomena and their relation to the critical exponents is a
promising and challenging question which deserves extensive future
investigation.

\vskip 0.3 cm We acknowledge financial support from NCET of M.O.E,
and NSF of China Project No. 10305002 and 60578014. We would also
like to thank Prof. F. X. Han for the stimulating discussions with
him.


\begin{thebibliography}{99}

\bibitem{nielsen00}  M. A. Nielsen and I. L. Chuang, Quantum computation and
quantum information (Cambridge University press, Cambridge, 2000).

\bibitem{sachdev99} S. Sachdev, quantum phase transition
(Cambridge University Press, Cambridge, 1999).
\bibitem{osborne02} T. J. Osborne, M. A. Nielsen, Phys. Rev. A
{\bf 66}, 032110(2002).

\bibitem{vidal03} G. Vidal, J. I. Latorre, E. Rico, and A. Kitaev,
Phys. Rev. Lett. {\bf 90}, 227902(2003).

\bibitem{osterloh02} A. Osterloh, L. Amico, G. Falci, R. Fazio,
Nature(London) {\bf 416}, 608(2002).

\bibitem{wu04} L. A. Wu, M. S. Sarandy, D. A. Lider, Phys. Rev.
Lett. {\bf 93}, 250404(2004).

\bibitem{li04} Y. Chen, P. Zanardi, Z. D. Wang, and F. C. Zhang,
e-print:quant-ph/0407228; Y. Chen, Z. D. Wang, and F. C. Zhang,
quant-ph/0512143.

\bibitem{gu04} S. J. Gu, S. S. Deng, Y. Q. Li, and H. Q. Lin,  Phys. Rev. Lett. {\bf 93},
086402(2004).

\bibitem{verstraete04} F. Verstraete, M. Popp, and J. I. Cirac,  Phys. Rev. Lett.
{\bf 92}, 027901(2004).

\bibitem{plenio99} M. B. Plenio, S. F. Huelga, A. Beige, and P. L. Knight,  Phys. Rev. A {\bf
59}, 2468(1999); M. B. Plenio, S. F. Huelga,  Phys. Rev. Lett.
{\bf 88}, 197901(2002).

\bibitem{bose99} S. Bose, P. L. Knight, M. B. Plenio, and V. Vedral,  Phys. Rev. Lett. {\bf 83},
5158(1999).

\bibitem{beige00} A. Beige, S. Bose, D. Braun, S. F. Huelga, P. L. Knight, M. B. Plenio, and V. Vedral,
 J. Mod. Opt. {\bf 47}, 2583(2000).

\bibitem{horodecki01} P. Horodecki, Phys. Rev. A {\bf 63},
022108(2001).

\bibitem{braun02} D. Braun, Phys. Rev. Lett. {\bf 89},
277901(2002).

\bibitem{yi03} X. X. Yi, C. S. Yu, L. Zhou, and H. S. Song,  Phys. Rev. A {\bf 68},
052304(2003).

\bibitem{beuatti03} F. Beuatti, R. Floreanini, M. Piani, Phys.
Rev. Lett. {\bf 91}, 070402(2003).

\bibitem{shresta05} S. Shresta, C. Anastopoulos, A. Dragulescu,
and B. L. Hu, Phys. Rev. A {\bf 71}, 022109(2005).

\bibitem{duan03} L. M. Duan, E. Demler, and M. D. Lukin, Phys.
Rev. Lett. {\bf 91}, 090402(2003).

\bibitem{porras04} D. Porras and J. I. Cirac, Phys. Rev. Lett.
{\bf 93}, 207901 (2004); X. L. Deng, D. Porras, and J. I. Cirac,
e-print quant-ph/0509197.

\bibitem{pachos04} J. K. Pachos and M. B. Plenio, Phys. Rev. Lett.
{\bf 93}, 056402(2004).

\bibitem{lieb61} E. Lieb, T. Schultz, D. Mattis, Annals of Physics
{\bf 16}, 407(1961).

\bibitem{katsura62} S. Katsura, Phys. Rev. {\bf 127}, 1058(1962).

\bibitem{note1} As addressed in \cite{lieb61}, $M$ is not
invarient under the Bogoliubov transformation, however its
evenness or oddness does. So that $e^{i\pi M}$ is invarient. The
terms with $e^{i\pi M}=-1$ give zero contribution to
$\Gamma_{ij;mn}(t)$, because the boundary terms disappear in this
case. For $e^{i\pi M}=1$, the boundary terms make changes of order
$1/N$ in $\cos(\theta_{ij,k}/2)$ and $\sin(\theta_{ij,k}/2)$.
Therefore they make corrections upto  order $1/N$ in
$\Gamma_{ij;mn}(t)$, which can be safely neglected in the limit
$N\rightarrow \infty$. Intutively, with $N\rightarrow \infty$, the
two cases of open boundary and cyclic boundary become the same, so
the boundary effect would disappear in this sense.

\bibitem{barouch70} E. Barouch and B. M. McCoy, Phys. Rev. A {\bf
2}, 1075(1970); Phys. Rev. A {\bf 3}, 786{1971}.

\bibitem{yang05} M. F. Yang, Phys. Rev. A {\bf 71},030302(2005).

\bibitem{zanardi05} P. Zanardi, N. Paunkovic, e-print:
quant-ph/0512249.

\bibitem{dutta01} A. Dutta, J. K. Bhattacharjee, Phys. Rev. B {\bf
64}, 184106(2001).

\bibitem{angelo05} Angelo C. M. Carollo and Jiannis K. Pachos,
Phys. Rev. Lett. {\bf 95}, 157203(2005).



\bibitem{hartmann05} M. J. Hartmann, M. E. Reuter, and M. B.
Plenio, New J. Phys. {\bf 8}, 94{2006}; see also,
e-print:quant-ph/0511185.

\bibitem{quan05}H. T. Quan, Z. Song, X. F. Liu, P. Zanardi, and C.
P. Sun, Phys. Rev. Lett. {\bf 96}, 140604(2006).

\end{thebibliography}
\end{document}